\documentclass[aps,prl,reprint,twocolumn,nofootinbib]{revtex4-2}

\usepackage{epsfig,amssymb,amsmath,physics,enumitem,qcircuit,bm}

\usepackage{ifthen}
\newboolean{suppmat}
\setboolean{suppmat}{true}
\ifthenelse{\boolean{suppmat}}{
    \newcommand{\appref}[2]{\hyperref[#2]{Appendix~#1}}
}{
    \newcommand{\appref}[2]{the Supplemental Material}
}

\usepackage{color,subfig}
\graphicspath{{./fig/}}

\usepackage{graphicx,ragged2e}
\usepackage{etoolbox}
\makeatletter
\renewcommand{\@makecaption}[2]{%
  \vskip\abovecaptionskip
  \sbox\@tempboxa{#1: #2}%
  \ifdim \wd\@tempboxa > \hsize
    \begin{center} 
      \small\justifying #1: #2
    \end{center}%
  \else
    \begin{center} 
      \small #1: #2
    \end{center}%
  \fi
  \vskip\belowcaptionskip}
\makeatother

\definecolor{myurlcolor}{rgb}{0,0,0.7}
\definecolor{myrefcolor}{rgb}{0.8,0,0}
\usepackage[unicode=true,pdfusetitle, bookmarks=true,bookmarksnumbered=false,bookmarksopen=false, breaklinks=false,pdfborder={0 0 0},backref=false,colorlinks=true, linkcolor=myrefcolor,citecolor=myurlcolor,urlcolor=myurlcolor]{hyperref}

\newtheorem{theorem}{Theorem}

\newtheorem{lemma}{Lemma}

\def\equationautorefname~#1\null{%
  Eq.~(#1)\null
}
\def\figureautorefname~#1\null{%
  FIG.~#1\null
}
\def\subfigureautorefname~#1\null{%
  FIG.~#1\null
}

\renewcommand{\a}{\hat{a}}
\renewcommand{\b}{\hat{b}}

\newcommand{\D}{\mathcal{D}}
\renewcommand{\S}{\mathcal{S}}
\newcommand{\R}{\mathcal{R}}
\newcommand{\n}{\hat{n}}
\newcommand{\q}{\hat{q}}
\newcommand{\p}{\hat{p}}

\renewcommand{\H}{\hat{H}}

\newcommand{\inv}{^{-1}}

\newcommand{\hft}{\frac{\theta}{2}}
\newcommand{\tq}{\tilde{q}}
\newcommand{\tp}{\tilde{p}}

\newcommand{\hf}[1]{\frac{#1}{2}}
\newcommand{\bma}{\bm{\alpha}}

\newcommand{\bmmu}{\bm{\mu}}
\newcommand{\bmsg}{\bm{\Sigma}}
\newcommand{\ketq}[1]{\ket{#1}_{q}}
\newcommand{\ketp}[1]{\ket{#1}_{p}}
\newcommand{\keta}[1]{\ket{#1}_{c}}

\newcommand{\braa}[1]{{}_{c}\!\bra{#1}}
\newcommand{\dyadcc}[1]{\ket{#1}_{c\,c}\!\bra{#1}}

\newcommand{\pf}[1]{\textit{Proof of \autoref{#1}.}}
\newcommand{\qed}{\hfill$\square$}
\newcommand{\sectext}[1]{\textit{#1}.---}

\newcommand{\appmomentumsampling}{A}
\newcommand{\appQGT}{B}
\newcommand{\appbeamsplitter}{C}
\newcommand{\appdisplacement}{D}
\newcommand{\appstatetransfer}{E}

\newcommand{\thetitle}{Quantum Coherent State Transform on Continuous-Variable Systems}
\newcommand{\theauthor}{
    Xi Lu$^{1,2}$,
    Bojko N. Bakalov$^{3}$ and
    Yuan Liu$^{2,4,5}$
}
\newcommand{\theaffiliation}{
    $^1$School of Mathematical Science, Zhejiang University, Hangzhou, 310027, China\\
    $^2$Department of Electrical and Computer Engineering, North Carolina State University, Raleigh, NC 27606, USA\\
    $^3$Department of Mathematics, North Carolina State University, Raleigh, NC 27695, USA\\
    $^4$Department of Computer Science, North Carolina State University, Raleigh, NC 27606, USA \\
    $^5$Department of Physics, North Carolina State University, Raleigh, NC 27606, USA
}

\begin{document}

\title{\thetitle}
\author{\theauthor}
\affiliation{\theaffiliation}
\begin{abstract}
    While continuous-variable (CV) quantum systems are believed to be more efficient for quantum sensing and metrology than their discrete-variable (DV) counterparts due to the infinite spectrum of their native operators, our toolkit of manipulating CV systems is still limited.
    We introduce the quantum coherent state transform~(QCST) and a framework for implementing it in CV quantum systems with two ancilla CV states and six two-mode SUM gates.
    Measurement of the resulting quantum state under the momentum eigenbasis is equivalent to a positive operator-valued measure (POVM) with elements $\{\frac{1}{\pi}\dyad{\alpha}\}_{\alpha\in\mathbb{C}}$, which provides an efficient way to learn the original CV state.
    Our protocol makes it possible to estimate the coherent state parameter within minimum-uncertainty precision using a single copy of the state, which finds applications in single-shot gate calibration of beam splitter and rotation gates to arbitrary precision.
    With repeated runs of our protocol, one can also estimate the parameters of any Gaussian state, which helps to calibrate other Gaussian gates, such as squeezing.
    For non-Gaussian states, our protocols can be used to perform Husimi Q-function tomography efficiently.
    With DV systems as ancilla instead, we can realize QCST approximately, which can be used to transfer CV states to DV states and back.
    The simplicity and broad applicability of the quantum coherent state transform make it an essential tool in continuous-variable quantum information science and engineering.
\end{abstract}
\maketitle

Quantum computing with hybrid continuous-variable~(CV, oscillator) and discrete-variable~(DV, qubit) systems has attracted much attention recently~\cite{liu2024hybrid,crane2024hybrid}, due to its ability to maintain a balance between easy-to-calibrate DV systems and resource-efficient CV systems.
While it could be hard to have universal control in pure CV systems, the assistance of DV systems can help in many tasks, such as generating non-Gaussian states~\cite{jacobs2007engineering,teoh2023dual}, generating universal instruction sets~\cite{eickbusch2022fast,job2023efficient,liu2021constructing}, error correction~\cite{grimsmo2021quantum,gottesman2001encoding}, quantum signal processing~\cite{low2017optimal,martyn2021grand,sinanan2024single}, quantum simulation of bosonic and fermionic systems~\cite{crane2024hybrid,kang2023leveraging}, etc.

A key feature of using CV quantum systems is their efficiency in sensing and metrology~\cite{giovannetti2004quantum,abbott2016observation,sinanan2024single}, which aim to extract classical information by manipulating quantum entanglement and measurement.
In general, the Heisenberg limit~(HL) can be achieved in estimating quantum parameters, i.e., $\Delta\varphi\propto N^{-1}$ using $N$ probes to estimate a parameter $\varphi$ encoded in a quantum process. This provides a quadratic improvement over the standard quantum limit~(SQL) $\Delta\varphi\propto N^{-1/2}$. 
Furthermore, the infinite spectrum of native CV operators makes it possible to achieve arbitrary precision in single-shot decision making~\cite{sinanan2024single}, which is crucial in some situations where the underlying signal occurs rarely, such as gravitational wave detection~\cite{abbott2016observation}.

Unique features of CV system metrology include homodyne/heterodyne detection~\cite{wang2020efficient,eichler2011experimental} and photon number detection~\cite{hann2018robust,johnson2010quantum}, which are fundamental tools in CV state learning tasks like CV state tomography~\cite{landon2018quantitative,he2024efficient,vogel1989determination,ourjoumtsev2006quantum}.
Due to the non-commutable nature of position and momentum in CV systems, the efficiency of those tools to learn CV states are limited.
A full tomography of general CV states is extremely inefficient, but tomography of Gaussian states is efficient~\cite{mele2024learning}.

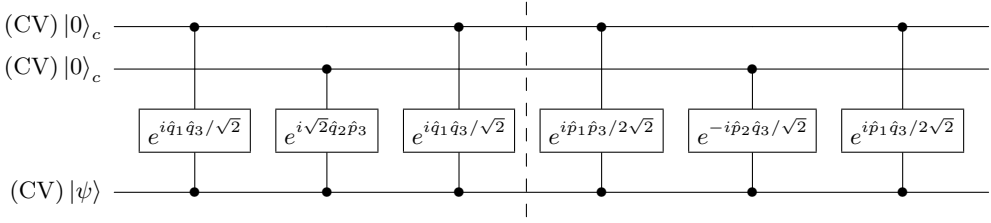
\begin{figure*}
    \centering
    $$
    \Qcircuit @C=1em @R=1.5em {
        \lstick{(\mathrm{CV})\keta{0}} & \ctrl{2} & \qw      & \ctrl{2} \barrier[-1cm]{3} & \ctrl{2} & \qw      & \ctrl{2} & \qw \\
        \lstick{(\mathrm{CV})\keta{0}} & \qw      & \ctrl{1} & \qw      & \qw      & \ctrl{1} & \qw      & \qw \\
        &
            \push{\boxed{e^{i\q_1\q_3/\sqrt{2}}}} &
            \push{\boxed{e^{i\sqrt{2} \q_2\p_3}}} &
            \push{\boxed{e^{i\q_1\q_3/\sqrt{2}}}} &
            \push{\boxed{e^{i\p_1\p_3/2\sqrt{2}}}} &
            \push{\boxed{e^{-i\p_2\q_3/\sqrt{2}}}} &
            \push{\boxed{e^{i\p_1\q_3/2\sqrt{2}}}} & \\
    \lstick{(\mathrm{CV})\ket{\psi}}    & \ctrl{-1} & \ctrl{-1} & \ctrl{-1} & \ctrl{-1} & \ctrl{-1} & \ctrl{-1} & \qw
    }
    $$
    \caption{The quantum circuit for QCST, where the subscripts of those quadratures indicate the index of the oscillator. Each gate is a two-mode SUM gate~\cite{liu2024hybrid,tzitrin2020progress}. For the purpose of Husimi Q-function sampling, one can just apply the first three gates, since the last three gates are used to reset the third CV state and does not change the measurement probabilities.}
    \label{fig:circuit-QCST}
\end{figure*}

In this Letter, we propose a novel protocol on CV systems called the \emph{quantum coherent state transform} (QCST), as well as a simple quantum circuit implementation, to transfer the non-commutable position and momentum information of one CV state into the commutable momentum information of two CV ancilla states.
Measurement of the two CV states in the momentum eigenbasis is equivalent to a POVM with elements $\{\frac{1}{\pi}\dyad{\alpha}\}_{\alpha\in\mathbb{C}}$, which helps us to learn position and momentum information of the original CV state at the same time.
The measurement result $\alpha$ obeys the distribution equal to the Husimi Q-function $Q(\alpha)$.
Given a single copy of an unknown coherent state, our protocol can estimate its parameter with minimum-uncertainty precision.
This makes it possible to perform single-shot gate calibration on more gates, such as beam splitter and rotation gates, compared to the existing single-shot binary decision making for displacement gates~\cite{sinanan2024single}.
For general Gaussian states, our protocol can estimate the Gaussian parameters from the first and second moments of the Q-function samples directly, which can help to calibrate other Gaussian gates such as single-mode squeezing.
Beyond Gaussian states, by running the protocol multiple times, one can reconstruct the Husimi Q-function of a general CV state, which does not require pointwise sampling on the phase space, and thus fits for large-region tomography.
Finally, by using DV systems as ancilla instead of CV systems, we can realize QCST approximately, which can be used for CV-DV state transfer.

\sectext{Quantum Coherent State Transform and Husimi Q-function Sampling}
We use the convention $\a=\frac{\q+i\p}{\sqrt{2}}$, for the annihilation operator $\a$, position operator $\q$ and momentum operator $\p$.
We use $\ketq{\cdot},\ketp{\cdot},\keta{\cdot},\ket{\cdot}_F$ to indicate position eigenstate, momentum eigenstate, coherent state and Fock state, respectively.

We define the \emph{quantum coherent state transform}~(QCST) of a CV state $\ket{\psi}$ as,
\begin{equation}
    QCST(\ket{\psi}) = \frac{2}{\sqrt{\pi}} 
    \iint
        {}_{c} \!\braket{\alpha}{\psi}
        \ketp{2\Re\alpha} \ketp{2\Im\alpha}
    \dd^2\alpha,
\label{eq:def-QCST}
\end{equation}
which encodes position and momentum information of a CV state $\ket{\psi}$ into the momentum amplitude of two CV states.

By measuring the first two CV states in the momentum eignebasis and obtaining $p_1,p_2$, we find that $\alpha:=\hf{p_1+ip_2}$ has the probability density function~(PDF) equal to the Husimi Q-function of $\ket{\psi}$,
\begin{equation}
    Q(\alpha) := \frac{1}{\pi} {}_{c} \!\bra{\alpha} \hat{\rho} \keta{\alpha},
\end{equation}
which is real non-negative on $\alpha\in\mathbb{C}$ and satisfies $\int Q(\alpha) \dd^2\alpha=1$.
One way to do that on hybrid CV-DV systems is given in \appref{\appmomentumsampling}{app:mom-sampl}.
We call this protocol \textit{Husimi Q-function sampling}~(HQS).
Our main result that realizes QCST is as follows.

\begin{theorem}[Quantum Coherent State Transform]\label{thm:husimi-q}
    The quantum circuit in \autoref{fig:circuit-QCST} implements the transform,
    \begin{equation}
        \keta{0} \keta{0} \ket{\psi} \mapsto QCST(\ket{\psi}) \keta{0}.
    \end{equation}
\end{theorem}

\pf{thm:husimi-q}
The unitary transformation by all three gates is,
\begin{equation}
    e^{i\q_1\q_3/\sqrt{2}}
    e^{i\sqrt{2} \q_2\p_3}
    e^{i\q_1\q_3/\sqrt{2}}
    =
    e^{i\sqrt{2} (\q_1\q_3+\q_2\p_3)}.
\end{equation}
The vacuum state can be written in the position basis as,
\begin{equation}
    \keta{0} = \pi^{-1/4} \int e^{-\hf{1}q^2} \ketq{q} \dd q.
\end{equation}
We find that the quantum state after the first three gates is,
\begin{equation}
\begin{aligned}
    &
    e^{i\sqrt{2} (\q_1\q_3+\q_2\p_3)} \keta{0} \keta{0} \ket{\psi}
    \\ = &
    \frac{1}{\sqrt{\pi}}
    \iint e^{-\hf{1}(q_1^2+q_2^2)}
    \ketq{q_1} \ketq{q_2} e^{i\sqrt{2} (q_1\q_3+q_2\p_3)} \ket{\psi}
    \dd q_1 \dd q_2
    \\ = &
    \frac{1}{2\sqrt{\pi^3}}
    \iiiint e^{-\hf{1}(q_1^2+q_2^2)-i(q_1p_1+q_2p_2)}
    \ketp{p_1} \ketp{p_2} 
    \\ & \times
    e^{i\sqrt{2} (q_1\q_3+q_2\p_3)} \ket{\psi}
    \dd q_1 \dd q_2 \dd p_1 \dd p_2
    \\ = &
    \frac{1}{2\sqrt{\pi}}
    \iint \ketp{p_1} \ketp{p_2} \dyadcc{\hf{p_1+ip_2}} \ket{\psi} \dd p_1 \dd p_2
    \\ = &
    \frac{2}{\sqrt{\pi}} 
    \iint
        {}_{c} \!\braket{\alpha}{\psi}
        \ketp{2\Re\alpha}
        \ketp{2\Im\alpha}
        \keta{\alpha}
    \dd^2\alpha.
\end{aligned}
\label{eq:first-3-gates}
\end{equation}
Here we used the identity,
\begin{equation}
    \dyadcc{\alpha} = \frac{1}{\pi} \iint e^{-\hf{1}(p^2+q^2)-2i(p\Re\alpha+q\Im\alpha)} e^{i\sqrt{2}(p\q+q\p)} \dd q \dd p,
\end{equation}
which is obtained by Wigner--Weyl transform from the Wigner function $W_{\alpha}(\beta) = \frac{2}{\pi} e^{-2|\beta-\alpha|^2}$.

The last three two-mode SUM gates give the unitary transform $e^{i(-\p_2\q_3+\p_1\p_3)/\sqrt{2}}$, which transforms \autoref{eq:first-3-gates} into,
\begin{equation}
\begin{aligned}
    &
    \frac{2}{\sqrt{\pi}} 
    \iint
        {}_{c} \!\braket{\alpha}{\psi}
        \ketp{2\Re\alpha}
        \ketp{2\Im\alpha}
    \\ & \times
        e^{-i\sqrt{2}(\Im\alpha\q_3-\Re\alpha\p_3)}
        \keta{\alpha}
    \dd^2\alpha
    \\ = &
    \frac{2}{\sqrt{\pi}} 
    \iint
        {}_{c} \!\braket{\alpha}{\psi}
        \ketp{2\Re\alpha}
        \ketp{2\Im\alpha}
        \keta{0}
    \dd^2\alpha.
\end{aligned}
\end{equation}
\qed

Since the last three gates in \autoref{fig:circuit-QCST} are used to reset the third state and do not change the measurement probabilities in the momentum eigenbasis, for the purpose of HQS, one may just apply the first half of \autoref{fig:circuit-QCST} followed by a measurement in the momentum eigenbasis.

\sectext{Generalization}
The results of HQS extend directly to the case where $\ket{\psi}$ is replaced by a mixed state $\rho$ due to the linearity of quantum circuits, and to the multimode case where one adds two ancilla oscillators. In the latter case, one applies the circuit for each of the $n$ modes and measures all ancilla oscillators to obtain $2 n$ momentum results whose probability density is equal to the joint Husimi Q-function.

QCST can also be generalized to \emph{Quantum Gaussian Transform} with the role of coherent state replaced by general Gaussian state, as discussed in \appref{\appQGT}{app:QGT}.

\sectext{Single-Shot Coherent State Estimation and Gate Calibration}
When $\ket{\psi}$ is the coherent state $\keta{\beta}$, the measurement has a PDF 
$
    Q_{\beta}(\alpha) = \frac{1}{\pi} e^{-\abs{\beta-\alpha}^2}
$.
If we use the measurement result $\alpha$ as an estimation of $\beta$, then
$
(\Delta\Re(\alpha-\beta))^2 = (\Delta\Im(\alpha-\beta))^2 = \frac{1}{2},
$
which achieves the minimum-uncertainty precision.

As an application, we show how coherent state parameter estimation can be used to estimate with only one query the parameters $\theta$ and $\phi$ in a beam splitter gate,
\begin{equation}
	U(\theta, \phi) = e^{-i\hft(e^{i\phi}\a^\dag \b + e^{-i\phi}\b^\dag \a)}.
\end{equation}
The beam splitter acts as $
U(\theta, \phi) \keta{\alpha} \keta{\beta}
=
\keta{\alpha'} \keta{\beta'}
$ where,
\begin{equation}
	\alpha'=\alpha\cos\hft + i\beta \sin\hft e^{i\phi},
    \quad
	\beta'=\beta\cos\hft + i\alpha \sin\hft e^{-i\phi}.
\end{equation}

Using known $\alpha$ and $\beta$ and estimation of the two output coherent state parameters, we can estimate $\theta$ and $\phi$ by finding the closest pair
\begin{equation}
	\left(
        \cos\hft, \sin\hft e^{i\phi}
    \right) \approx \left(
        \frac{\alpha'\alpha^* + \beta\beta'^*}{|\alpha|^2 + |\beta|^2}, i\frac{\alpha\beta'^* - \alpha'\beta^*}{|\alpha|^2 + |\beta|^2}
    \right).
\end{equation}

Note that the standard deviation of $\alpha',\beta'$ is constant.
If we choose $\alpha=\beta>0$ in the initial state, then the estimation error is $\Delta\theta=\mathcal{O}(\alpha^{-1})$, $\Delta\phi=\mathcal{O}(\alpha^{-1}\theta^{-1})$, which shows that we can achieve arbitrarily high precision by using initial coherent states far from the vacuum state.
This achieves the Heisenberg limit in terms of the resource of the initial state, as we discuss in \appref{\appbeamsplitter}{app:beam-splitter}.

Similarly, HQS can also be used to calibrate the phase-space rotation gate $\R(\theta)=e^{-i\theta\n}$ in a single shot by estimating $\keta{\alpha'}=\R(\theta)\keta{\alpha}$ with precision $\mathcal{O}(\abs{\alpha}^{-1})$.
However, for the task of displacement gate calibration, a similar idea of estimating $\alpha'$ in $\keta{\alpha'}=\D(\beta)\keta{\alpha}$ in a single shot gives constant precision instead of arbitrary precision as for rotation and beam splitter gates.
Fortunately, we can achieve arbitrary precision with a slightly different protocol in \appref{\appdisplacement}{app:disp}.

\sectext{Gaussian State Estimation}
The Husimi Q-function of any pure Gaussian state is also Gaussian,
\begin{equation}
    Q(\alpha) = \frac{1}{2\pi\sqrt{\mathrm{det}(\bmsg)}} \exp\left(
        -\hf{1}
        (\bma-\bmmu)^\top
        \bmsg\inv
        (\bma-\bmmu)
    \right),
    \label{eq:gauss-q}
\end{equation}
where $\bma = \begin{bmatrix} \Re\alpha \\ \Im\alpha \end{bmatrix}$,
$\bmmu$ and $\bmsg$ are the mean and the covariance matrix of $\bma$.
Given a CV state that is promised to be a pure Gaussian state with unknown parameters, one can reconstruct the parameters in \autoref{eq:gauss-q} easily from $M$ samples $\{\alpha_j\}$ by, 
\begin{equation}
    \tilde{\bmmu} = \frac{1}{M} \sum_{j=1}^{M} \bma_j
    \text{ and }
    \tilde{\bmsg} = \frac{1}{M-1} \sum_{j=1}^{M} (\bma_j-\tilde\bmmu)(\bma_j-\tilde\bmmu)^\top,
    \label{eq:est-gauss}
\end{equation}
where $\bma_j = \begin{bmatrix} \Re\alpha_j \\ \Im\alpha_j \end{bmatrix}$.
The error of the mean and covariance matrices scale as $\mathcal{O}(M^{-1/2})$.
Similar discussion on sampling complexity for multi-mode case can be found in \cite{mele2024learning}.

\begin{figure}
    \centering
    \includegraphics[width=\linewidth]{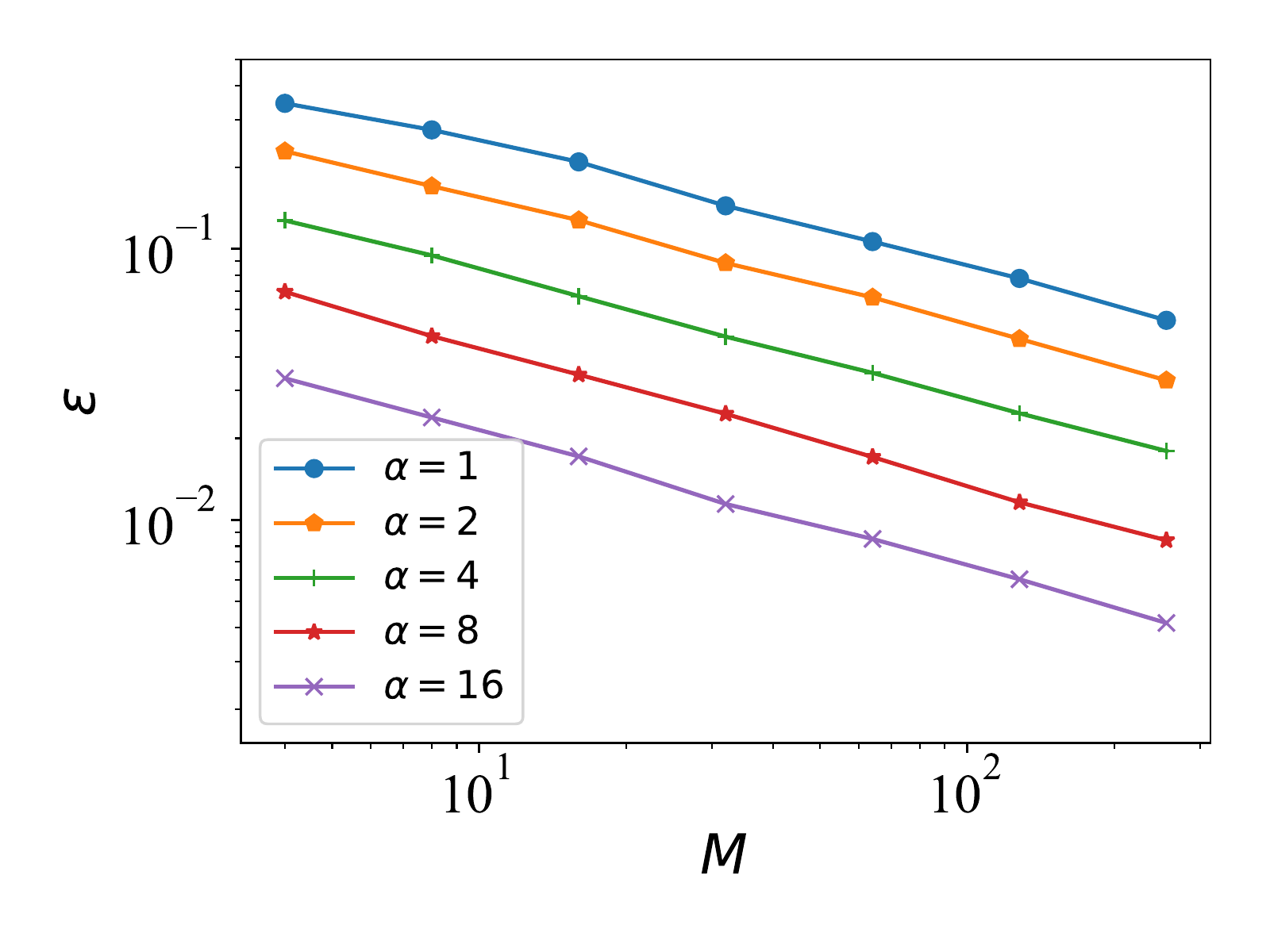}
    \caption{The error $\epsilon=\sqrt{\mathbb{E}[|\tilde{\xi}-\xi|^2]}$ of squeezing parameter estimation, which shows approximately $\epsilon\sim M^{-1/2}|\alpha|^{-1}$. For each configuration we repeat 1000 times to calculate the error.}
    \label{fig:squeeze-est}
\end{figure}

\begin{figure*}
    \centering
    \includegraphics[width=\linewidth]{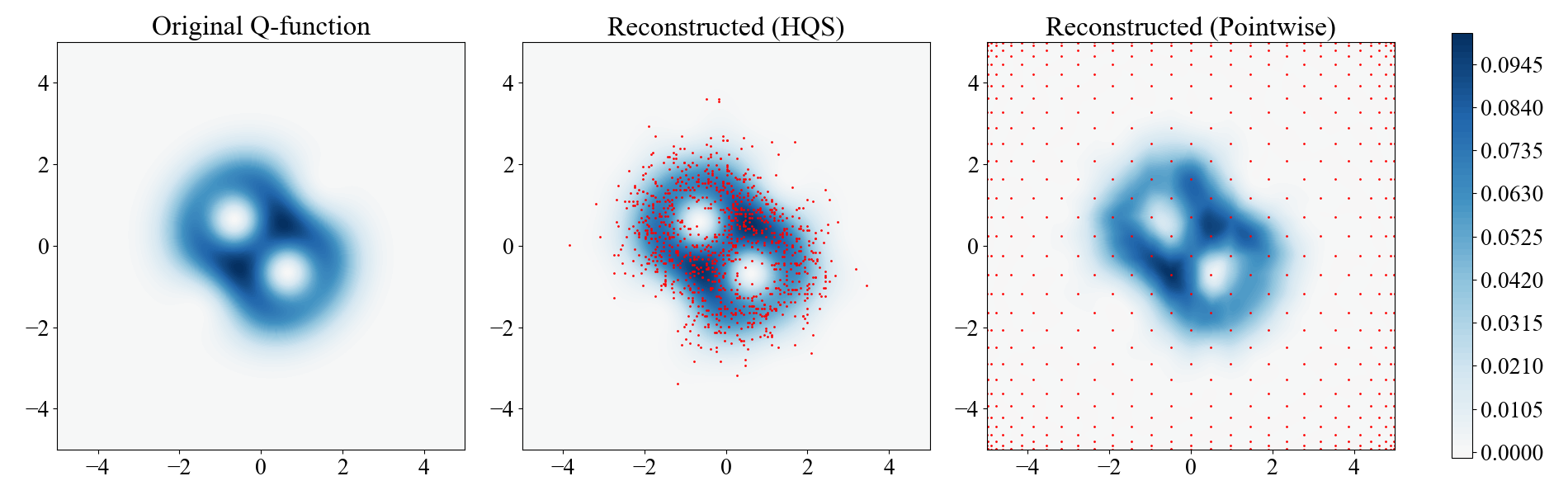}
    \caption{Numerical experiment results of Q-function tomography on the test states $\frac{\ket{0}_F+\ket{4}_F}{2}+\frac{i}{\sqrt{2}}\ket{2}_F$. In the middle, we show 1024 samples in total shown as red dots, and use $\Gamma=32$ in MLE. On the right, we show the results of pointwise method with 561 Padua points and 1024 samples on each point.}
    \label{fig:qtomo-eight}
\end{figure*}

\begin{figure}
    \centering
    \includegraphics[width=\linewidth]{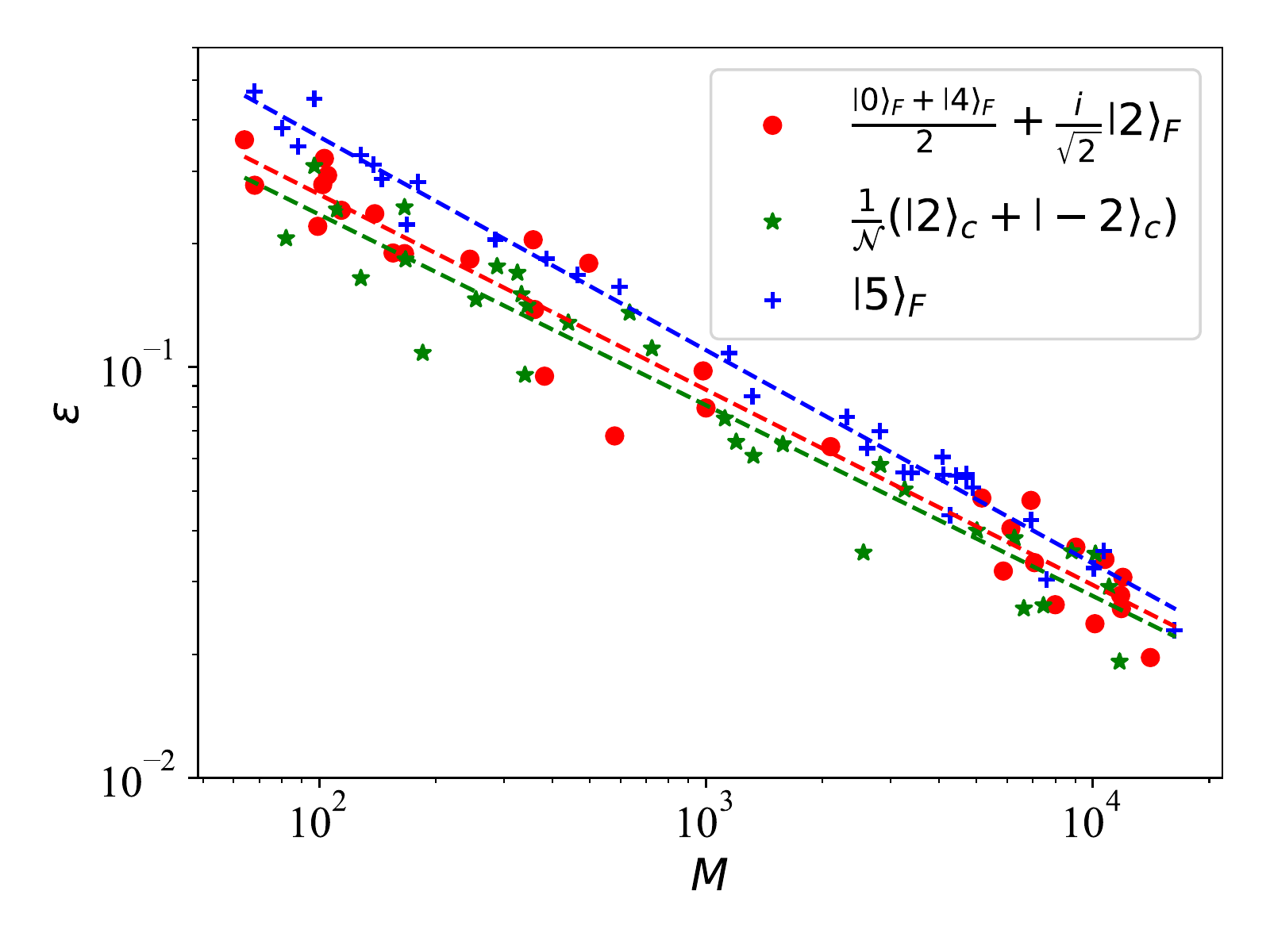}
    \caption{Error scaling for three test states using our method. Each shows a $\epsilon\sim M^{-1/2}$ scaling.}
    \label{fig:qtomo-error}
\end{figure}

As an application, we show how to calibrate a single-mode squeezing gate with our protocol.
Applying a squeezing gate $\S(\xi)=e^{\hf{1}(\xi^\dag\a^2-\xi\a^{\dag 2})}$ to a coherent initial state $\keta{\alpha}$ $(\alpha>0)$, the output state has Husimi Q-function given by \autoref{eq:gauss-q} with,
\begin{equation}
    \bmmu = R_{\theta} \begin{bmatrix}
        \alpha e^{-r} \cos\hft \\
        -\alpha e^{r} \sin\hft
    \end{bmatrix},
    \quad
    \bmsg = R_{\theta} \begin{bmatrix}
        \frac{1+e^{-2r}}{4} & 0 \\
        0 & \frac{1+e^{2r}}{4}
    \end{bmatrix} R_{\theta}\inv,
    \label{eq:gauss-xi}
\end{equation}
where $\xi=re^{i\theta}$ and $R_{\theta}=\begin{bmatrix}
    \cos\hft & -\sin\hft \\
    \sin\hft &  \cos\hft
\end{bmatrix}$.
With $M$ samples, we first estimate $\tilde{\bmmu}$ and $\tilde{\bmsg}$ with \autoref{eq:est-gauss} and use \autoref{eq:gauss-q} to reconstruct the Q-function $\tilde{Q}(\alpha)$, then numerically search for $\tilde{\xi}$ that minimizes a simple loss function $\norm{\bmmu(\xi)-\tilde{\bmmu}}+\norm{\bmsg(\xi)-\tilde{\bmsg}}_F$, with $\bmmu(\xi)$ and $\bmsg(\xi)$ given in \autoref{eq:gauss-xi} and the subscript $F$ denoting the Frobenius norm.
We perform a numerical experiment for different $M,\alpha$ and random $\xi\in\{\xi\in\mathbb{C}:|\xi|\le 1\}$, and show the results in \autoref{fig:squeeze-est}.

\sectext{Q-function Tomography}
If we run HQS enough times, the resulting distribution will converge to the Husimi Q-function.
This can help us reconstruct the Husimi Q-function of a general CV state.
We study the case of pure state tomography, in which the Husimi Q-function of a general pure state $
    \sum_{k=0}^{\infty} \psi_k \ket{k}_{F}
$ is,
\begin{equation}
    Q(\alpha) = \frac{1}{\pi} e^{-\abs{\alpha}^2} \abs{\sum_{k=0}^{\infty}\frac{\psi_k^*}{\sqrt{k!}}\alpha^k}^2.
    \label{eq:para-q}
\end{equation}

To reconstruct the Husimi Q-function from a finite number of random samples $\{\alpha_j\}_{j=1}^{M}$ with PDF equal to $Q(\alpha)$, one can assume a Fock level cutoff $\Gamma$, i.e., all $\psi_k=0$ for $k\ge\Gamma$ in \autoref{eq:para-q}, then use \textit{Maximum Likelihood Estimation}~(MLE) and minimize
\begin{equation}
    -\frac{1}{M} \sum_{j=1}^{M} \log Q(\alpha;\{\psi_j\}),
    \text{ such that }
    \sum_{k=0}^{\Gamma-1} |\psi_k|^2 = 1.
    \label{eq:mle-opt}
\end{equation}

We compare this method with the existing pointwise method for Q-function tomography~\cite{landon2018quantitative}, which measures $|\braket{\psi}{\alpha}|^2$ at some specific points and then uses Lagrange interpolation to recover a continuous Q-function.
At each point, a single measurement result returns one bit of information, so one needs to run multiple times at each Padua points.
We show the results of the numerical experiment in \autoref{fig:qtomo-eight}.
Our HQS protocol followed by MLE behaves better even with the number of samples equal to existing protocols for a single Padua point.
A major advantage of our protocol is that our samples can find where the state is in a large range of phase space by themselves, while in pointwise methods most samples return very little information since the Q-function values are nearly zero in most of the area.
A more quantitative analysis of the error scaling of our method is shown in \autoref{fig:qtomo-error}, where the error is defined by the difference $\epsilon=\iint\abs{Q(\alpha)-\tilde{Q}(\alpha)}\dd^2\alpha$ between the original Q-function $Q(\alpha)$ and the reconstructed one $\tilde{Q}(\alpha)$.
The results show an error scaling of $\mathcal{O}(M^{-1/2})$, which is a typical error scaling of MLE, since the Fisher information grows linearly with the number of independently and identically distributed samples.

\sectext{CV/DV State Transfer}
It is possible to approximate QCST using only DV ancilla systems and conditional displacement gates.
The key idea is to discretize the ancilla CV states in \autoref{fig:circuit-QCST} into linear combinations of position eigenstates $\sum_{j=0}^{N-1} c_j \ketq{q_j}$, where $q_j=(j-\hf{N-1})\lambda$ and $\lambda>0$ is the grid size.
Then two-mode SUM gates can be simulated by a sequence of conditional displacement gates in hybrid CV-DV systems like,
\begin{eqnarray}
    e^{i\q_1(\alpha\a_3^\dagger-\alpha^*\a_3)}
    &\mapsto&
    \sum_j \dyad{j} \otimes \D_3(q_j \alpha),
    \label{eq:dv-map-1}
    \\
    e^{i\p_1(\alpha\a_3^\dagger-\alpha^*\a_3)}
    &\mapsto&
    \sum_j \dyad{\tilde{j}} \otimes \D_3(p_j \alpha),
    \label{eq:dv-map-2}
\end{eqnarray}
where $\ket{\tilde{j}} = \frac{1}{\sqrt{N}} \sum_{k=0}^{N-1} e^{i2\pi jk/N} \ket{k}$ is the $j$-th Fourier basis state and $p_j := \frac{\pi}{N\lambda}\left[\left((j+\hf{N})\bmod N\right) - \hf{N}\right]$.
We refer to the details in \appref{\appstatetransfer}{app:cd-only}.

As $\lambda,\frac{1}{N\lambda}\to 0$ simultaneously, the ancilla DV initial state approaches the CV vacuum state, and the transfer approaches the perfect QCST, in which the CV state is reset to the vacuum state and all the information is transferred to the ancilla systems.
This is known as \textit{CV-to-DV state transfer}~\cite{hastrup2022universal,liu2024toward}.
Its inverse gives \textit{DV-to-CV state transfer}, which requires a CV vacuum state and a Husimi-encoded DV state as input and outputs the original CV state approximately.
The CV state $\ket{\psi}$ is encoded as $\ip{\alpha}{\psi}$ on a discrete lattice of $\alpha$, which is a different approach from the position wave function encoding in \cite{hastrup2022universal}.
An advantage of our DV-to-CV transfer is that the CV state is initialized to the vacuum state, instead of an nonphysical state that can only be prepared approximately in \cite{hastrup2022universal}.

\sectext{Conclusion}
We introduce the Quantum Coherent State Transform in \autoref{thm:husimi-q}, which is used to extract information about the position and momentum of a CV quantum state simultaneously.
We show its use in many applications: (1) single-shot coherent state estimation and single-shot gate calibration of beam splitter and rotation gates with Heisenberg error scaling; (2) Gaussian state estimation, which can be used to calibrate more Gaussian gates like single-shot squeezing gate; (3) Q-function tomography for general CV states; (4) CV-DV state transfer. The latter uses DV ancilla states instead of CV states in \autoref{thm:husimi-q} to transfer CV states to DV states and reset CV states to the vacuum state. DV states can be transferred back to CV states by inverting the protocol on encoded DV states and a vacuum CV state.

Our protocol serves as a new tool to extract information from CV quantum systems and to interact CV systems with DV systems.
The new form of encoding a CV state into the momentum amplitude of two CV states or DV systems could inspire new techniques in quantum error correction and new ways to learn CV states in quantum simulation.
Our work could provide the fundamental toolkit in scalable CV technologies and help bridge the quantum and classical realms across quantum science and general physics.

\bibliographystyle{unsrt}
\bibliography{ref.bib}

\ifthenelse{\boolean{suppmat}}{
    
\clearpage
\onecolumngrid
\begin{center}
	\textbf{\large Supplemental Material for ``\thetitle"}
	\vspace{0.5cm}
	\\\theauthor
	\\\textit{\theaffiliation}
\end{center}

\appendix
\setcounter{page}{1}
\thispagestyle{empty}
\setcounter{figure}{0}

\section{\appmomentumsampling. Momentum Eigenbasis Measurement Using Conditional Displacement}\label{app:mom-sampl}
\setcounter{equation}{0}
\renewcommand{\theequation}{\appmomentumsampling\arabic{equation}}

\begin{lemma}[Momentum Eigenbasis Measurement]\label{lem:q-sampling}
	The measurement probability of the quantum circuit
	$$
		\Qcircuit @C=1em @R=0.7em {
			\lstick{(\mathrm{DV})\sum_{j=0}^{N-1} c_{j} \ket{j}} & \multigate{1}{ACD^{(N,-\lambda)}} & \gate{QFT^\dagger} & \meter{} & \cw{/} & \cw \\
			\lstick{(\mathrm{CV})\ket{\psi}}             & \ghost{ACD^{(N,-\lambda)}}        & \qw                & \qw & \qw    & \qw
		}
	$$
	in which the \textit{Array Controlled Displacement}~(ACD) is defined as,
	\begin{equation}
        \mathrm{ACD}^{(N,\alpha)} := \sum_j \dyad{j} \otimes \D\left(
            \alpha\left(j-\frac{N-1}{2}\right)
        \right),
	\end{equation}
	and $QFT^\dagger$ is the inverse quantum Fourier transform, is given by,
	\begin{equation}
        Pr(j) = \frac{1}{N}
        \int \left|
    		\sum_{j'=0}^{N-1} c_{j'} e^{-2\pi ij'(p-\frac{2\pi j}{N\lambda})}
    	\right|^2 \abs{\psi(p)}^2 \dd p,
	\end{equation}
	where $\psi(p)$ is the momentum-basis wave function of the oscillator.
\end{lemma}

Note that the $\mathrm{ACD}^{(N,\alpha)}$ can be constructed using conditional displacement gates $\D_c(\alpha):=e^{-\sigma_z(\alpha\a^\dag-\alpha^*\a)}$ as follows,
$$
\Qcircuit @C=2em @R=1em {
    \lstick{} & \ctrl{4} & \qw    &\qw       & \qw      & \qw \\
    \lstick{} &          & \cdots &          &          &     \\
    \lstick{} & \qw      & \qw    & \ctrl{2} & \qw      & \qw \\
    \lstick{} & \qw      & \qw    & \qw      & \ctrl{1} & \qw \\
    \lstick{} & 
        \push{\boxed{\D_c(2^{n-2} \alpha)}} & \cdots &
        \push{\boxed{\D_c(2^{0}   \alpha)}} &
        \push{\boxed{\D_c(2^{-1}  \alpha)}} & \\
    \lstick{} & \ctrl{-1} & \qw & \ctrl{-1} & \ctrl{-1} & \qw
}
$$

\pf{lem:q-sampling}
The state before measurement is,
\begin{equation}
\begin{aligned}
    &
    (QFT^\dag\otimes I) \sum_{j=0} c_j\ket{j} \D\left(i\lambda(j-\frac{N-1}{2})\right) \int \psi(q) \ket{q} \dd q
    \\ = &
    \frac{1}{\sqrt{N}} \sum_{j=0}^{N-1} \ket{j} \sum_{j'=0}^{N-1} c_{j'} e^{-2\pi jj'/N} \int e^{i\lambda p(j-\frac{N-1}{2})} \psi(p) \ketp{p} \dd p.
\end{aligned}
\end{equation}

Then the probability of measurement result $j$ is,
\begin{equation}
\begin{aligned}
    Pr(j) = &
    \frac{1}{N} \sum_{j',j''=0}^{N-1} c_{j'} c_{j''}^* e^{-2\pi j(j'-j'')/N} \int e^{i\lambda p(j'-j'')} \abs{\psi(p)}^2 \dd p
    \\ = &
    \frac{1}{N}
    \int \left|
		\sum_{j'=0}^{N-1} c_{j'} e^{-2\pi ij'(\lambda p-\frac{2\pi j}{N})}
	\right|^2 \abs{\psi(p)}^2 \dd p,
\end{aligned}
\label{eq:discrete-pj}
\end{equation}
which proves the lemma.
\qed

\begin{figure}
    \centering
    \includegraphics[width=.6\linewidth]{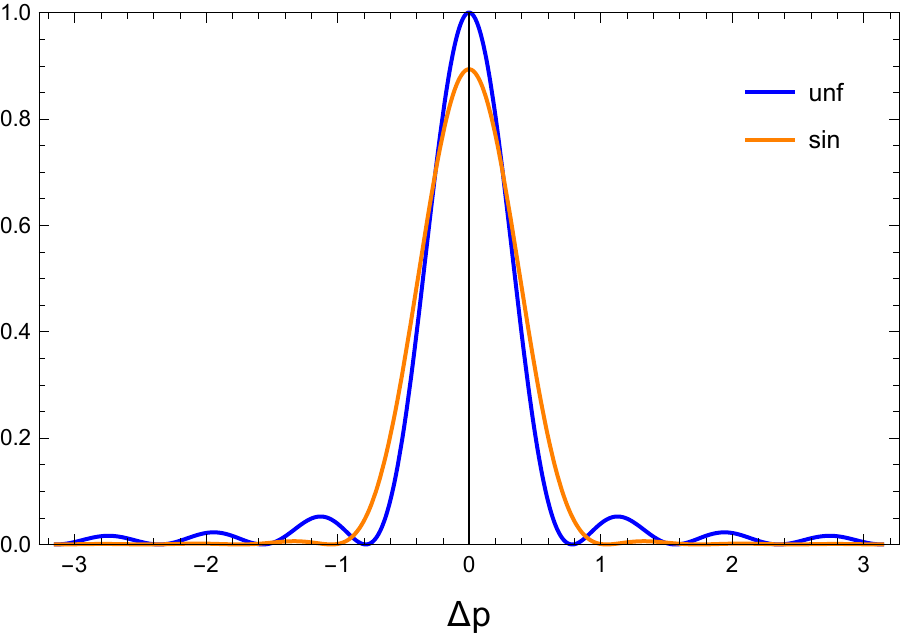}
    \caption{The function $\frac{1}{N}\left|
        \sum_{j'=0}^{N-1} c_{j'} e^{-2\pi ij'\Delta p}
    \right|^2$ with respect to $\Delta p\in[-\pi,\pi]$, when $N=8$.}
    \label{fig:unf-vs-sin}
\end{figure}

In standard QPE one chooses the uniform initial state $c_j=\frac{1}{\sqrt{N}}$ (denoted as \textit{unf}), while a more robust choice in quantum metrology is the sine initial state $c_j=\sqrt{\frac{2}{N+1}}\sin\left(\frac{j+1}{N+1}\pi\right)$ (denoted as \textit{sin}).
We plot the function $\left|
    \sum_{j'=0}^{N-1} c_{j'} e^{-ij'\Delta p}
\right|^2$ against $\Delta p$ in \autoref{fig:unf-vs-sin}.
The \textit{unf} initial state is more concentrated around $\Delta p=0$, but also has a long tail, while the \textit{sin} initial state is not as concentrated around zero but decays very quickly.

Note that,
\begin{equation}
    \int \left|
        \sum_{j'=0}^{N-1} c_{j'} e^{-ij'\lambda(p-\tp)}
    \right|^2 \dd p
    =
    \frac{2\pi}{\lambda} \sum_{j'=0}^{N-1} |c_{j'}|^2
    =
    \frac{2\pi}{\lambda},
\end{equation}
always holds, where the integral is over any length-$\frac{2\pi}{\lambda}$ period.
If we take $N\to\infty$, since the integrand goes to zero at any $p\notin\tp+\frac{2\pi}{\lambda}\mathbb{Z}$ for either \textit{unf} or \textit{sin} initial state,
\begin{equation}
    \left|
        \sum_{j'=0}^{N-1} c_{j'} e^{-ij'\lambda(p-\tp)}
    \right|^2 \to \frac{2\pi}{\lambda} \sum_{p'\in\tp+\frac{2\pi}{\lambda}\mathbb{Z}} \delta(p-p'),
\end{equation}
and the estimation $\tp$ becomes a continuous variable with probability density function~(PDF),
\begin{equation}
    Pr(j) = \frac{2\pi}{N\lambda} \sum_{p'\in\tp+\frac{2\pi}{\lambda}\mathbb{Z}} |\psi(p')|^2
    \quad \Rightarrow \quad
    Pr(\tq) = \sum_{p'\in\tp+\frac{2\pi}{\lambda}\mathbb{Z}} |\psi(p')|^2.
\end{equation}
Finally, taking $\lambda\to 0$, we obtain sampling from the PDF $|\psi(p)|^2$.

In real implementations of this protocol, we should choose $\lambda$ small enough such that $\psi(q)$ almost lies in one period $[-\frac{\pi}{\lambda},\frac{\pi}{\lambda}]$ and thus $\sum_{p'\in\tp+\frac{2\pi}{\lambda}\mathbb{Z}} |\psi(p')|^2 \approx |\psi(\tp)|^2$, and $N$ large enough so that in \autoref{eq:discrete-pj},
\begin{equation}
    Pr(j) \approx \frac{2\pi}{N\lambda}\abs{
        \psi\left(
            \frac{2\pi}{N\lambda}
            \left(\left((j+\hf{N})\bmod N\right) - \hf{N}\right)
        \right)
    }^2.
\end{equation}
In a similar fashion, one can do position eigenbasis measurement, or even measurement from an arbitrary angle $(\cos\phi)\q+(\sin\phi)\p$.

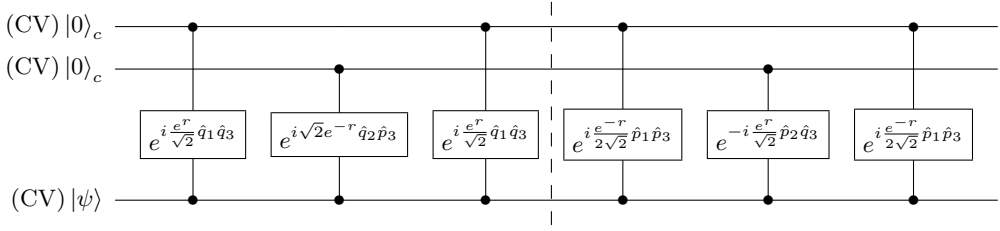
\begin{figure*}
    \centering
    $$
    \Qcircuit @C=1em @R=1.5em {
        \lstick{(\mathrm{CV})\keta{0}} & \ctrl{2} & \qw      & \ctrl{2} \barrier[-.95cm]{3} & \ctrl{2} & \qw      & \ctrl{2} & \qw \\
        \lstick{(\mathrm{CV})\keta{0}} & \qw      & \ctrl{1} & \qw      & \qw      & \ctrl{1} & \qw      & \qw \\
        &
            \push{\boxed{e^{i  \frac{e^{r}}{\sqrt{2}}   \q_1\q_3}}} &
            \push{\boxed{e^{i  \sqrt{2} e^{-r}          \q_2\p_3}}} &
            \push{\boxed{e^{i  \frac{e^{r}}{\sqrt{2}}   \q_1\q_3}}} &
            \push{\boxed{e^{i  \frac{e^{-r}}{2\sqrt{2}} \p_1\p_3}}} &
            \push{\boxed{e^{-i \frac{e^{r}}{\sqrt{2}}   \p_2\q_3}}} &
            \push{\boxed{e^{i  \frac{e^{-r}}{2\sqrt{2}} \p_1\p_3}}} & \\
    \lstick{(\mathrm{CV})\ket{\psi}}    & \ctrl{-1} & \ctrl{-1} & \ctrl{-1} & \ctrl{-1} & \ctrl{-1} & \ctrl{-1} & \qw
    }
    $$
    \caption{The quantum circuit for QGT.}
    \label{fig:circuit-QGT}
\end{figure*}

\section{\appQGT. Generalization to Quantum Gaussian Transform}
\label{app:QGT}
\setcounter{equation}{0}
\renewcommand{\theequation}{\appQGT\arabic{equation}}

Our QCST can be generalized to \emph{Quantum Gaussian Transform}~(QGT), with coherent states replaced by general Gaussian states.
As an example, let $\ket{\alpha,r}=\S(r)\keta{\alpha}$ $(r\in\mathbb{R})$ be a squeezed coherent state with a squeezing parameter $r$, and define
\begin{equation}
    QGT_r(\ket{\psi}) = \frac{2}{\sqrt{\pi}} 
    \iint
        \braket{\alpha,r}{\psi}
        \ketp{2\Re\alpha} \ketp{2\Im\alpha}
    \dd^2\alpha.
\label{eq:def-QGT}
\end{equation}
The quantum circuit that performs,
\begin{equation}
    \keta{0} \keta{0} \ket{\psi} \mapsto QGT_r(\ket{\psi}) \ket{0,r}.
\end{equation}
is shown in \autoref{fig:circuit-QGT}.
Indeed, the quantum state after the first three gates is,
\begin{equation}
\begin{aligned}
    &
    e^{i\sqrt{2} (e^{r}\q_1\q_3+e^{-r}\q_2\p_3)} \keta{0} \keta{0} \ket{\psi}
    \\ = &
    \frac{1}{\sqrt{\pi}}
    \iint e^{-\hf{1}(q_1^2+q_2^2)}
    \ketq{q_1} \ketq{q_2} e^{i\sqrt{2} (q_1e^{r}\q_3+q_2e^{-r}\p_3)} \ket{\psi}
    \dd q_1 \dd q_2
    \\ = &
    \frac{1}{\sqrt{\pi^3}}
    \iiiint e^{-\hf{1}(q_1^2+q_2^2)-i(q_1p_1+q_2p_2)}
    \ketp{p_1} \ketp{p_2}
    e^{i\sqrt{2} (q_1e^{r}\q_3+q_2e^{-r}\p_3)} \ket{\psi}
    \dd q_1 \dd q_2 \dd p_1 \dd p_2
    \\ = &
    \frac{1}{\sqrt{\pi}}
    \iint \ketp{p_1} \ketp{p_2} \dyad{\hf{p_1+ip_2},r} \ket{\psi} \dd p_1 \dd p_2
    \\ = &
    \frac{2}{\sqrt{\pi}} 
    \iint
        \braket{\alpha,r}{\psi}
        \ketp{2\Re\alpha}
        \ketp{2\Im\alpha}
        \keta{\alpha}
    \dd^2\alpha.
\end{aligned}
\end{equation}
Here we used the identity,
\begin{equation}
    \dyad{\alpha,r} = \frac{1}{\pi} \iint e^{-\hf{1}(p^2+q^2)-2i(p\Re\alpha+q\Im\alpha)} e^{i\sqrt{2}(pe^{r}\q+qe^{-r}\p)} \dd q \dd p,
\end{equation}
which is obtained by Wigner--Weyl transform from its Wigner function,
\begin{equation}
    W_{\alpha,r}(\beta) = \frac{2}{\pi} e^{-2[(e^{r}\Re\beta-\Re\alpha)^2+(e^{-r}\Im\beta-\Im\alpha)^2]}.
\end{equation}
Similarly, the remaining three gates transforms the state into,
\begin{equation}
\begin{aligned}
    &
    \frac{2}{\sqrt{\pi}} 
    \iint
        \braket{\alpha,r}{\psi}
        \ketp{2\Re\alpha}
        \ketp{2\Im\alpha}
        e^{-i\sqrt{2}(e^{r}\Im\alpha\q_3-e^{-r}\Re\alpha\p_3)}
        \ket{\alpha,r}
    \dd^2\alpha
    \\ = &
    \frac{2}{\sqrt{\pi}} 
    \iint
        \braket{\alpha,r}{\psi}
        \ketp{2\Re\alpha}
        \ketp{2\Im\alpha}
        \ket{0,r}
    \dd^2\alpha.
\end{aligned}
\end{equation}

\begin{figure}
    \centering
    \hspace{2cm}
    \subfloat[The DV-ancilla quantum circuit for single-shot displacement gate calibration.]{
        $$
        \Qcircuit @C=1em @R=0.7em {
            \lstick{(\mathrm{DV})\sum_{j=0}^{N-1} c_{j} \ket{j}} & \qw{/} & \multigate{2}{LCD} & \qw               & \multigate{2}{LCD^{\dagger}} & \gate{QFT^\dagger} & \meter{} & \cw \\
            \lstick{(\mathrm{DV})\sum_{k=0}^{N-1} c_k \ket{k}}   & \qw{/} & \ghost{LCD}        & \qw               & \ghost{LCD^\dag}        & \gate{QFT^\dagger} & \meter{} & \cw \\
            \lstick{(\mathrm{CV})\ket{\psi}}                     & \qw    & \ghost{LCD}        & \gate{\D(\alpha)} & \ghost{LCD^\dag}        & \qw                & \qw & \qw
        }
        $$
        \label{fig:dv-ancilla-disp}
    }
    \quad
    \subfloat[The implementation of the $\mathrm{LCD}$ gate.]{
        $$
        \Qcircuit @C=1em @R=1em {
            \lstick{}    & \qw{/} & \ctrl{2} & \qw      & \ctrl{2} & \qw \\
            \lstick{}    & \qw{/} & \qw      & \ctrl{1} & \qw      & \qw \\
            \lstick{} & &
                \push{\boxed{ACD^{(N, \hf{\lambda})}}} &
                \push{\boxed{ACD^{(N,   -i\lambda )}}} &
                \push{\boxed{ACD^{(N, \hf{\lambda})}}} & \\
            \lstick{} & \qw & \ctrl{-1} & \ctrl{-1} & \ctrl{-1} & \qw
        }
        $$
    }
    \\
    \subfloat[The CV-ancilla quantum circuit for single-shot displacement gate calibration.]{
        $$
        \Qcircuit @C=1em @R=1.5em {
            \lstick{(\mathrm{CV})\keta{0}} & \ctrl{2} & \qw      & \ctrl{2} & \qw & \ctrl{2} & \qw      & \ctrl{2} & \qw \\
            \lstick{(\mathrm{CV})\keta{0}} & \qw      & \ctrl{1} & \qw      & \qw & \qw      & \ctrl{1} & \qw      & \qw \\
            &
                \push{\boxed{e^{i\lambda\q_1\q_3/2}}} &
                \push{\boxed{e^{i\lambda\q_2\p_3}}} &
                \push{\boxed{e^{i\lambda\q_1\q_3/2}}} & &
                \push{\boxed{e^{-i\lambda\p_1\p_3/2}}} &
                \push{\boxed{e^{-i\lambda\p_2\q_3}}} &
                \push{\boxed{e^{-i\lambda\p_1\q_3/2}}} & \\
        \lstick{(\mathrm{CV})\ket{\psi}}    & \ctrl{-1} & \ctrl{-1} & \ctrl{-1} & \gate{\D(\alpha)} & \ctrl{-1} & \ctrl{-1} & \ctrl{-1} & \qw
        }
        $$
        \label{fig:cv-ancilla-disp}
    }
    \caption{The quantum circuit for single-shot displacement gate calibration.}
    \label{fig:disp}
\end{figure}
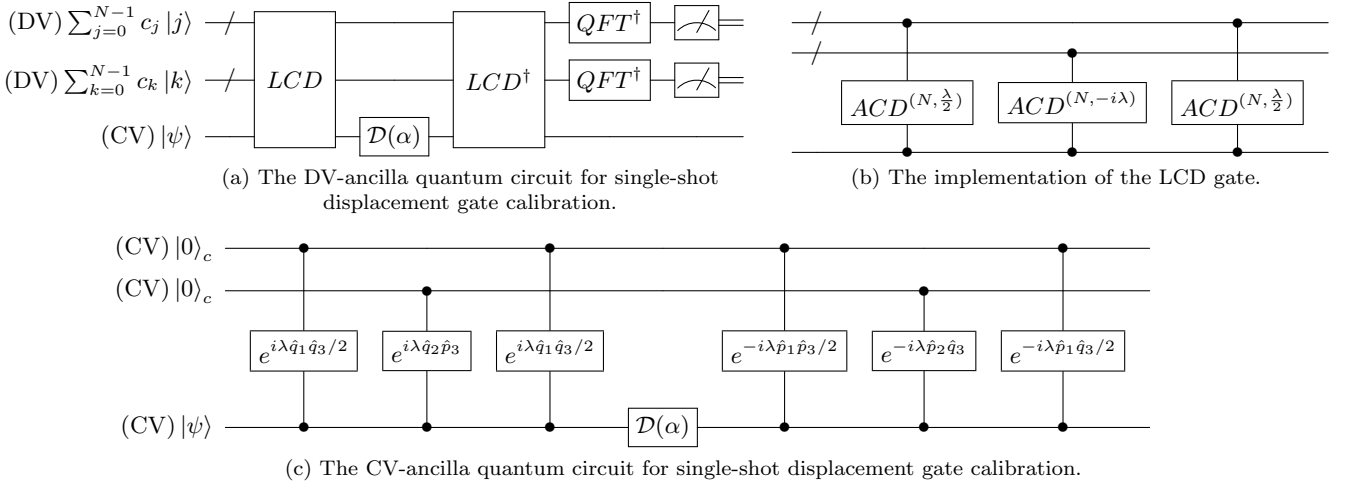

\section{\appbeamsplitter. Heisenberg Scaling of Single-Shot Beam Splitter Gate Calibration}
\label{app:beam-splitter}
\setcounter{equation}{0}
\renewcommand{\theequation}{\appbeamsplitter\arabic{equation}}

Quantum parameter estimation of $\theta$ in $\ket{\psi(\theta)} = e^{-i\theta\H}\ket{\psi}$ is said to achieve the Heisenberg limit if $\Delta\psi=\mathcal{O}(\Delta H^{-1})$, where $\Delta H = \sqrt{\ev{\H^2}{\psi}-\ev{\H}{\psi}^2}$.
In our single-shot beam splitter gate parameter estimation, $\ket{\psi}=\keta{\alpha}\keta{\alpha}$ and $\H=\hf{1}(e^{i\phi}\a^\dag\b+e^{-i\phi}\a\b^\dag)$.
Then,
\begin{equation}
    \ev{\H}{\psi} = |\alpha|^2 \cos\phi,
    \quad
    \ev{\H^2}{\psi} = |\alpha|^4 \cos^2\phi + \hf{|\alpha|^2},
\end{equation}
which gives $\Delta H=\frac{|\alpha|}{\sqrt{2}}$; hence, the estimation of $\theta$ achieves the Heisenberg limit.

\section{\appdisplacement. Single-Shot Displacement Gate Calibration}
\label{app:disp}
\setcounter{equation}{0}
\renewcommand{\theequation}{\appdisplacement\arabic{equation}}

The displacement parameter $\alpha$ in an unknown displacement operator $\D(\alpha)$ can also be estimated to arbitrary accuracy using a single shot. Our method for doing it is not based on coherent parameter estimation, but on quantum phase estimation.
The quantum circuit is shown in \autoref{fig:disp}, in which $\ket{\psi}$ can be arbitrary CV state, and the LCD gate is defined as,
\begin{equation}
    LCD =
    \sum_{j,k} \dyad{j} \otimes \dyad{k} \otimes \D\left(q_j+i q_k\right),
    \label{def:LCD}
\end{equation}
where $q_j=(j-\hf{N-1})\lambda$.

In \autoref{fig:dv-ancilla-disp}, the gates from LCD to its inverse perform the transformation,
\begin{equation}
\begin{aligned}
    LCD^\dag \D(\alpha) LCD
    = &
    \sum_{j,k} \dyad{j}\otimes\dyad{k}\otimes[\D(-q_j+i q_k)\D(\alpha)\D(q_j-i q_k)]
    \\ = &
    \left[
        \sum_{j,k} e^{
            i2(q_j\Im\alpha+q_k\Re\alpha)
        } \dyad{j}\otimes\dyad{k}
    \right]\otimes\D(\alpha),
\end{aligned}
\end{equation}
implying that the oscillator can be decoupled by the end, leaving a phase factor in the DV registers that can be extracted by 2-dimensional quantum phase estimation.
One can choose $N$ and $\lambda$ according to the parameter range and target precision.

The circuit in \autoref{fig:cv-ancilla-disp} performs the transformation,
\begin{equation}
\begin{aligned}
    &
    e^{-i\lambda(\q_1\q_3+\q_2\p_3)} e^{i\sqrt{2}(\Im\alpha\p_3-\Re\alpha\q_3)} e^{i\lambda(\q_1\q_3+\q_2\p_3)} \keta{0} \keta{0} \ket{\psi}
    \\ = &
    \frac{1}{\sqrt{\pi}} \iint e^{-\hf{1}(q_1^2+q_2^2)} e^{i\lambda(q_1\q_3+q_2\p_3)} e^{i\sqrt{2}(\Im\alpha\p_3-\Re\alpha\q_3)} e^{-i\lambda(q_1\q_3+q_2\p_3)} \ketq{q_1} \ketq{q_2} \ket{\psi} \dd q_1 \dd q_2
    \\ = &
    \frac{1}{\sqrt{\pi}} \iint e^{-\hf{1}(q_1^2+q_2^2)-i\sqrt{2}\lambda(q_1\Im\alpha+q_2\Re\alpha)} \ketq{q_1} \ketq{q_2}  \dd q_1 \dd q_2
    \otimes
    e^{i\sqrt{2}(\Im\alpha\p_3-\Re\alpha\q_3)} \ket{\psi}
    \\ = &
    \frac{1}{2\sqrt{\pi^3}}
    \iint e^{-\hf{1}q_1^2-iq_1(p_1+\sqrt{2}\lambda\Im\alpha)} \ketp{p_1} \dd q_1 \dd p_1
    \otimes
    \iint e^{-\hf{1}q_2^2-iq_2(p_2+\sqrt{2}\lambda\Re\alpha)} \ketp{p_2} \dd q_2 \dd p_2
    \otimes
    e^{i\sqrt{2}(\Im\alpha\p_3-\Re\alpha\q_3)} \ket{\psi}
    \\ = &
    \frac{1}{\sqrt{\pi}}
    \int e^{-\hf{1}(p_1+\sqrt{2}\lambda\Im\alpha)^2} \ketp{p_1} \dd p_1
    \otimes
    \int e^{-\hf{1}(p_2+\sqrt{2}\lambda\Re\alpha)^2} \ketp{p_2} \dd p_2
    \otimes
    e^{i\sqrt{2}(\Im\alpha\p_3-\Re\alpha\q_3)} \ket{\psi}.
\end{aligned}
\end{equation}

By measuring the momentum of the first two modes and obtaining $\tilde{p}_1,\tilde{p}_2$, our estimation to $\alpha$ is
$
    \tilde{\alpha} = -\frac{\tilde{p}_2+i\tilde{p}_1}{\sqrt{2}\lambda}
$.
The standard deviation error of $\Re\tilde{\alpha}$ and $\Im\tilde{\alpha}$ are both $\frac{1}{\sqrt{2}\lambda}$.
One can estimate $\alpha$ to arbitrary precision with a single shot by choosing large enough $\lambda$.

\begin{figure}
    \centering
    \subfloat[CV-ancilla QCST.]{
        $$
        \Qcircuit @C=1em @R=1.5em {
            \lstick{(\mathrm{CV})\keta{0}} & \ctrl{2} & \qw      & \ctrl{2} & \ctrl{2} & \qw      & \ctrl{2} & \qw \\
            \lstick{(\mathrm{CV})\keta{0}} & \qw      & \ctrl{1} & \qw      & \qw      & \ctrl{1} & \qw      & \qw \\
            &
                \push{\boxed{e^{i\q_1\q_3/\sqrt{2}}}} &
                \push{\boxed{e^{i\sqrt{2} \q_2\p_3}}} &
                \push{\boxed{e^{i\q_1\q_3/\sqrt{2}}}} &
                \push{\boxed{e^{i\p_1\p_3/2\sqrt{2}}}} &
                \push{\boxed{e^{-i\p_2\q_3/\sqrt{2}}}} &
                \push{\boxed{e^{i\p_1\q_3/2\sqrt{2}}}} & \\
        \lstick{(\mathrm{CV})\ket{\psi}}    & \ctrl{-1} & \ctrl{-1} & \ctrl{-1} & \ctrl{-1} & \ctrl{-1} & \ctrl{-1} & \qw
        }
        $$
        \label{fig:circuit-QCST-encoding-cv}
    }
    \\
    \subfloat[DV-ancilla approximate QCST.]{
        $$
        \Qcircuit @C=.7em @R=1em {
            \lstick{(\mathrm{DV})}    & \qw{/} & \ctrl{2} & \qw      & \ctrl{2} & \gate{QFT} & \ctrl{2} & \qw      & \ctrl{2} & \qw \\
            \lstick{(\mathrm{DV})}    & \qw{/} & \qw      & \ctrl{1} & \qw      & \gate{QFT} & \qw      & \ctrl{1} & \qw      & \qw \\
            \lstick{} & &
                \push{\boxed{ACD^{(N,\hf{i\lambda})}}} &
                \push{\boxed{ACD^{(N,    -\lambda )}}} &
                \push{\boxed{ACD^{(N,\hf{i\lambda})}}} & &
                \push{\boxed{ACD_*^{(N,-\frac{\pi}{2N\lambda})}}} &
                \push{\boxed{ACD_*^{(N,-i\frac{\pi}{N\lambda})}}} &
                \push{\boxed{ACD_*^{(N,-\frac{\pi}{2N\lambda})}}} & \\
            \lstick{(\mathrm{CV})\ket{\psi}} & \qw & \ctrl{-1} & \ctrl{-1} & \ctrl{-1} & \qw & \ctrl{-1} & \ctrl{-1} & \ctrl{-1} & \qw
        }
        $$
        \label{fig:circuit-QCST-encoding-dv}
    }
    \caption{Comparison of CV-ancilla QCST and DV-ancilla approximate QCST. }
    \label{fig:cd-only}
\end{figure}
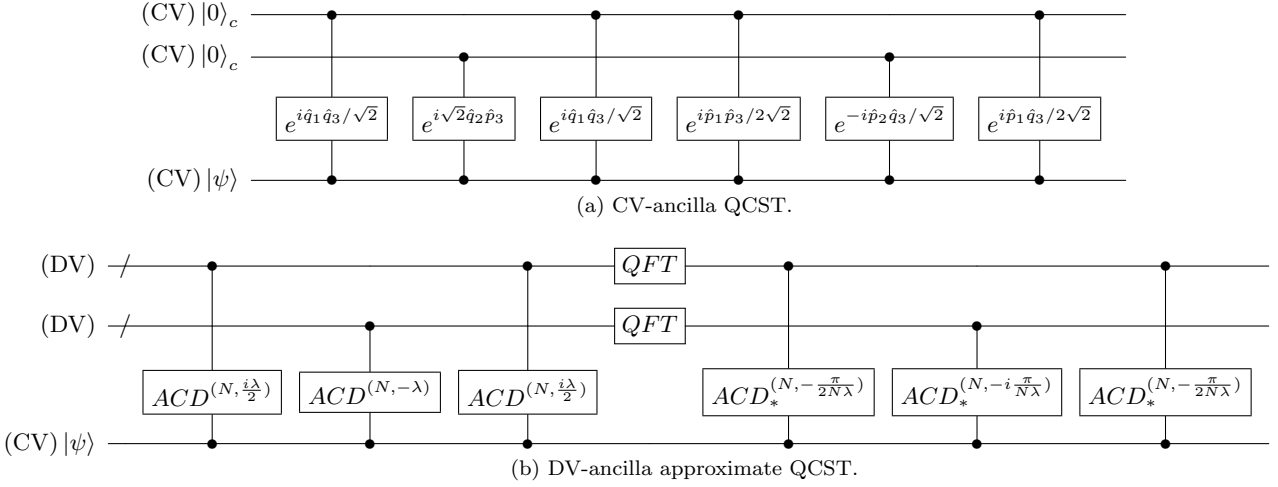

\section{\appstatetransfer. DV-ancilla Implementation of QCST and CV-DV State Transfer}
\label{app:cd-only}
\setcounter{equation}{0}
\renewcommand{\theequation}{\appstatetransfer\arabic{equation}}

\begin{figure}
    \centering
    \includegraphics[width=.95\linewidth]{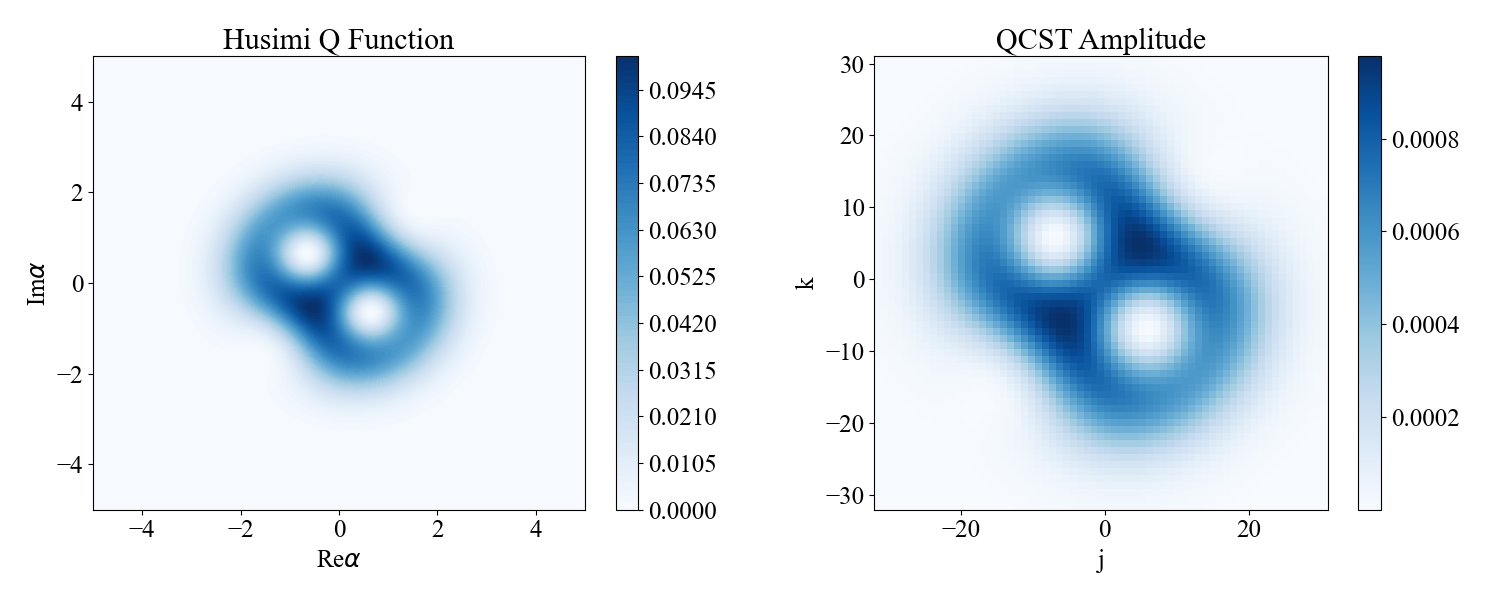}
    \caption{The Husimi Q-function and the QCST coefficient amplitude $\frac{1}{N^2}
    \abs{
        \braa{0} \sum_{j',k'=0}^{N-1} c_{j'} c_{k'} \D(q_{j'}+iq_{k'}) \D(p_j+ip_k) \ket{\psi}
    }^2$, where we move the indices $\hf{N},\cdots,N-1$ to $-\hf{N},\cdots,-1$ for better illustration, for the test state $\frac{\ket{0}_F+\ket{4}_F}{2}+\frac{i}{\sqrt{2}}\ket{2}_F$ and parameters $N=64,\lambda=0.5$ }
    \label{fig:circuit-QCST-dv}
\end{figure}

\begin{figure}
    \centering
    \includegraphics[width=.6\linewidth]{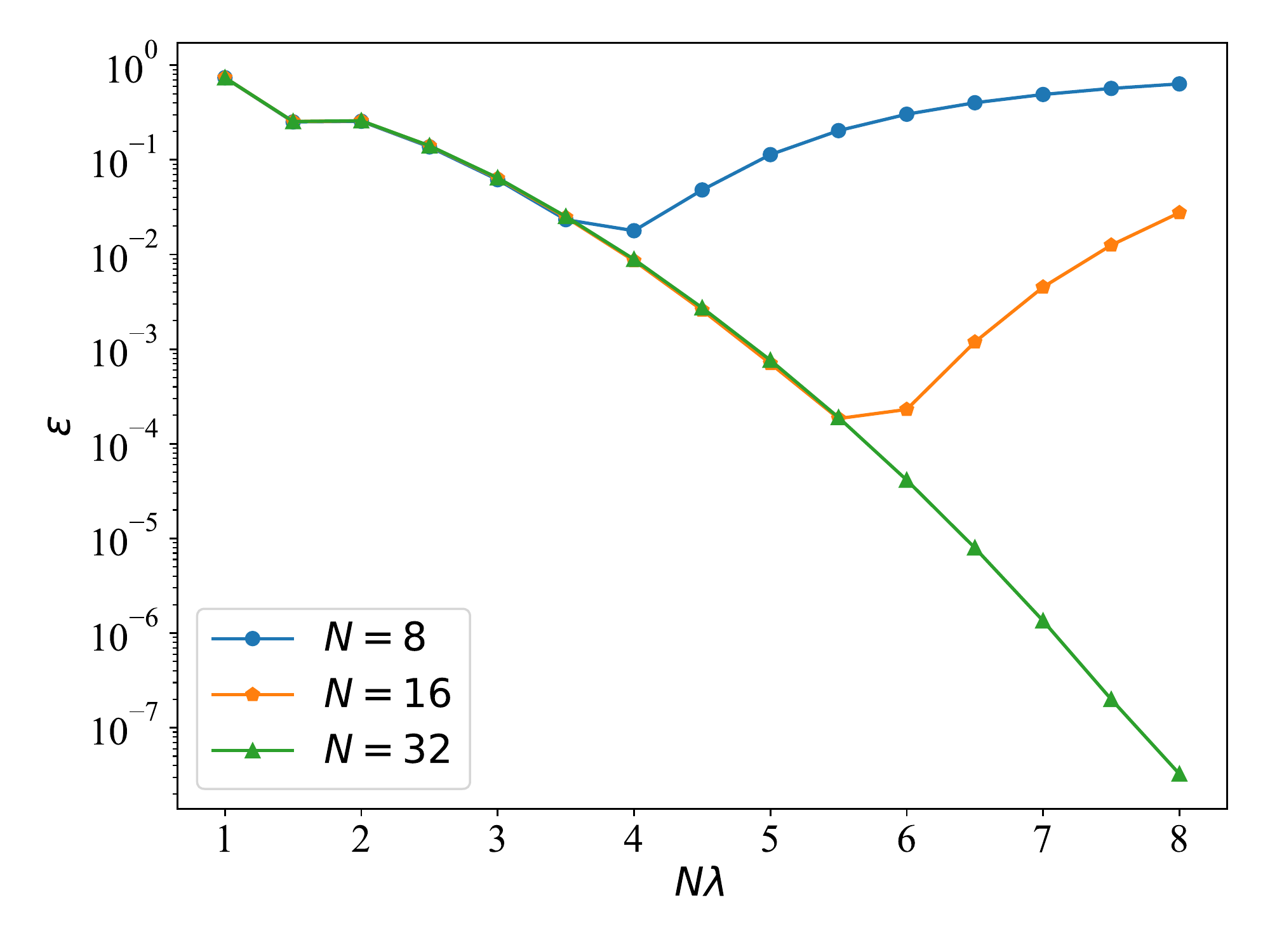}
    \caption{The error \autoref{eq:error-dv} for different $N$ and $\lambda$, for the test state $\frac{\ket{0}_F+\ket{4}_F}{2}+\frac{i}{\sqrt{2}}\ket{2}_F$.}
    \label{fig:error-dv}
\end{figure}

Given $N=2^n$ $(n\in\mathbb{Z}_+)$ and $\lambda>0$, one can use DV ancilla systems instead of CV to perform QCST \autoref{eq:def-QCST} in the main text approximately, as shown in \autoref{fig:cd-only}.
The $ACD_*^{(N,\alpha)}$ gate is defined as,
\begin{equation}
    \mathrm{ACD}_*^{(N,\alpha)} := \sum_j \dyad{j} \otimes \D\left(
        \alpha \left(\left((j+\hf{N})\bmod N\right) - \hf{N}\right)
    \right),
\end{equation}
which is similar to ACD gate but has different ordering.
We use the ACD gates in DV-ancilla circuit to simulate the two-mode SUM gate in the CV-ancilla circuit, according to the mappings \autoref{eq:dv-map-1} and \autoref{eq:dv-map-2}.

With $\sum_{j=0}^{N-1} c_j \ket{j}$ being the two DV initial states, the quantum circuit \autoref{fig:circuit-QCST-encoding-dv} goes as,
\begin{equation}
\begin{aligned}
    &
    \sum_{j,k=0}^{N-1} c_j c_k \ket{j}\ket{k} \ket{\psi}
    \\ \xrightarrow{\text{First Three Gates}} &
    \sum_{j,k=0}^{N-1} c_j c_k \ket{j}\ket{k} \D(-q_{k}+i q_{j}) \ket{\psi}
    \\ \xrightarrow{QFT} &
    \frac{1}{N} \sum_{j,k=0}^{N-1} \ket{j}\ket{k} \sum_{j',k'=0}^{N-1} c_{j'} c_{k'} e^{i 2\pi (jj'+kk')/N} \D(-q_{k'}+i q_{j'}) \ket{\psi}
    \\ \xrightarrow{\text{Remaining Gates}} &
    \frac{1}{N} \sum_{j,k=0}^{N-1} \ket{j}\ket{k} \sum_{j',k'=0}^{N-1} c_{j'} c_{k'} e^{i 2\pi (jj'+kk')/N} \D(p_j+ip_k) \D(-q_{k'}+iq_{j'}) \ket{\psi}
    \\ = &
    \frac{1}{N} \sum_{j,k=0}^{N-1} \ket{j}\ket{k} \sum_{j',k'=0}^{N-1} c_{j'} c_{k'} \D(q_{j'}+iq_{k'}) \D(p_j+ip_k) \ket{\psi},
\end{aligned}
\label{eq:husimi-encoding-dv}
\end{equation}
where $p_j := \frac{\pi}{N\lambda}\left[\left((j+\hf{N})\bmod N\right) - \hf{N}\right]$.

If one chooses $\{c_j\}$ so that
$
\sum_{j',k'=0}^{N-1} c_{j'} c_{k'} \D(q_{j'}+i q_{k'})
$
is approximately proportional to $\dyad{0}_{\alpha}$ in the limit as $N\to\infty$, $\lambda\to 0$, then the CV state is approximately reset to vacuum and almost all information is transferred to the DV systems.

To illustrate how the DV-ancilla QCST stores the information from the CV state, we plot the amplitude information of a test state and compare to its Husimi Q-function, as shown in \autoref{fig:circuit-QCST-dv}.
From \autoref{eq:husimi-encoding-dv}, the error of the protocol can be defined and calculated as,
\begin{equation}
\begin{aligned}
    \epsilon := &
    1 - \ev{(I_{DV}\otimes\dyad{0}_{CV})}{\tilde{\psi}}
    \\ = &
    1 - \frac{1}{N^2} \sum_{j,k=0}^{N-1} \abs{
        \braa{0} \sum_{j',k'=0}^{N-1} c_{j'} c_{k'} \D(q_{j'}+iq_{k'}) \D(p_j+ip_k) \ket{\psi}
    }^2.
\end{aligned}
\label{eq:error-dv}
\end{equation}

In the results of the numerical experiments shown in \autoref{fig:error-dv}, we see that with sufficiently large $N$, the error decreases exponentially with respect to $N\lambda$, which defines the gap between two adjacent $p_j$. One needs a large enough $N$ to keep that exponential decrease, as the discrete QCST only captures information of the Husimi Q-function in a finite square (see \autoref{fig:circuit-QCST-dv}). One would need a large enough $N$ to make sure that the most important part of the Husimi Q-function is included in that square, while failing to do so makes the error grow back again.

}{}

\end{document}